\documentclass{amsart}
\usepackage{amsmath}
\usepackage[mathcal]{eucal}
\usepackage{array}
\newcolumntype{L}{>{\centering\arraybackslash}m{3cm}}
\newtheorem{theorem}{Theorem}[section]
\newtheorem{lemma}[theorem]{Lemma}

\theoremstyle{definition}
\newtheorem{definition}[theorem]{Definition}
\newtheorem{example}[theorem]{Example}

\theoremstyle{remark}

\numberwithin{equation}{section}

\begin{document}
	
	\title{Reversible and Reversible Complement Cyclic codes over a class of non-chain rings}
	
	
	\author{Nikita Jain} 
	\address{Punjab Engineering College (Deemed to be University), Chandigarh}
	\email{nikitajain.phd19appsc@pec.edu.in}
	\thanks{ Nikita Jain would like to thank Council of Scientific and Industrial Research (CSIR) India, for providing fellowship in support of this research.
	}
	
	\author{Sucheta Dutt}
	\address{Punjab Engineering College (Deemed to be University), Chandigarh}
	\email{sucheta@pec.edu.in}
	
	\author{Ranjeet Sehmi}
	\address{Punjab Engineering College (Deemed to be University), Chandigarh}
	\email{rsehmi@pec.edu.in}
	
	
	
	\begin{abstract}
	In this paper, necessary and sufficient conditions for a cyclic code of arbitrary length over the non-chain rings $Z_{4}+\nu Z_{4}$ for $\nu^{2} \in \{0,1,\nu,2\nu,3\nu,2+\nu,2+3\nu,3+2\nu\}$ to be a reversible cyclic code have been established. Also, conditions for a cyclic code over these non-chain rings to be a reversible complement cyclic code which are necessary as well as sufficient have been determined. Some examples of reversible and reversible complement cyclic codes over these rings have also been presented.
	\end{abstract}
	\maketitle
	\section{Introduction}
	In algebraic coding theory, the class of cyclic codes is one of the important classes of codes. Cyclic codes have been extensively studied over rings after the remarkable work done by Calderbank et al. \cite{Calderbank} in which a Gray map has been introduced to show that some non- linear binary codes can be viewed as binary images of linear codes over $Z_{4}$.

The class of reversible codes is one of the useful classes of codes due to their role in DNA computing and reterival systems. Reversible codes over fields were first introduced by J.Massey \cite{massey}. A Necessary and sufficient condition for a cyclic code of odd length over $Z_{4}$ to be a reversible cyclic code  has been obtained by Abualrub and Siap \cite{reversible over $Z_{4}$}. The reversiblity conditions for a cyclic code of length $n$ relatively prime to $p$ over $Z_{p^k}$ have been obtained by H.Islam and O.Parkash \cite{$Z_p^{k}$}. Reversible cyclic codes over Galois rings have been studied by J.Kaur et al.\cite{jasbir thesis}. The structure of reversible cyclic codes of arbitrary length over the finite chain ring $F_{4}+\nu F_{4}, \nu^{2}=0$ has been obtained by Srinivasulu and Bhaintwal \cite{bhaintwal}. The structure of reversible cyclic codes of arbitrary length over the ring $F_{4}+\nu F_{4}+\nu^{2} F_{4}, \nu^{3}=0$ has been determined by J. Liu and H. Liu \cite{Jie Liu}. The conditions for a cyclic code to be a reversible cyclic code of arbitrary length over $F_{q}+\nu F_{q}+\cdots+\nu^{k-1}F_{q}, \nu^{k}=0$ and $k \geq 2$ have been obtained by O.Parkash et al. \cite{some chain ring,class of chain rings}. The necessary and sufficient conditions for a cyclic code to be a reversible cyclic code of odd length over the non-chain ring $Z_{4}+\nu Z_{4}, \nu^{2}=0$ have been established by S. Pattanayak, A.Kumar \cite{basepaper} and over the non-chain ring $Z_{4}+\nu Z_{4}, \nu^{2}=1$ have been obtained by H. Dinh et al. \cite{$u^{2}=1$}.

The class of reversible complement cyclic codes have also been extensively studied by many researchers due to their rich applications in DNA based computations. A vast literature is available on reversible complement cyclic codes over  different rings \cite{rev comp1,rev comp 2,rev comp 3,jasbir DAM}. Necessary and sufficient conditions for a cyclic code to be a reversible complement cyclic code over Galois rings have been obtained by J.Kaur et al. \cite{jasbir DAM}. A necessary and sufficient condition for a cyclic code to be a reversible complement cyclic code of odd length over the non-chain ring $Z_{4}+\nu Z_{4}, \nu^{2}=0$ has been established by S. Pattanayak, A.Kumar \cite{basepaper} and over the non-chain ring $Z_{4}+\nu Z_{4}, \nu^{2}=1$ has been obtained by H. Dinh et al. \cite{$u^{2}=1$}. 

The manuscript is organised as follows: In section 2, some basic definitions have been recalled. In section 3, sufficient and necessary conditions  for a cyclic code to be a reversible cyclic code over the rings $Z_{4}+\nu Z_{4}$ for $\nu^{2} \in \{0,\nu,2\nu,3\nu,1,3+2\nu,2+\nu,2+3\nu\}$ have been obtained. In section 4, conditions which are necessary as well as sufficient for a cyclic code to be a reversible complement cyclic code have been determined over the rings $Z_{4}+\nu Z_{4}$ for $\nu^{2} \in \{0,\nu,2\nu,3\nu,1,3+2\nu,2+\nu,2+3\nu\}$. 
 
 \section{Preliminaries}
Let $\mathtt{R}$ be a finite commutative ring with unity. If all the ideals of a ring $\mathtt{R}$ form a chain under the inclusion relation, then $\mathtt{R}$ is a chain ring. If not, then $\mathtt{R}$ is a non-chain ring.
A linear code $\mathtt{C}$ of length $n$ over the ring $\mathtt{R}$ is a $\mathtt{R}$- submodule of $\mathtt{R}^{n}$ and its elements are known as codewords of $\mathtt{C}.$ For a codeword $(\mathtt{s}_{_{0}},\mathtt{s}_{_{1}},\cdots,\mathtt{s}_{_{n-1}}) \in \mathtt{C},$ if $(\mathtt{s}_{_{n-1}},\mathtt{s}_{_{0}},\cdots,\mathtt{s}_{_{n-2}}) \in \mathtt{C},$ then $\mathtt{C}$ is said to be a cyclic code of length $n$ over $\mathtt{R}.$ A codeword $\mathtt{s}=(\mathtt{s}_{_{0}},\mathtt{s}_{_{1}},\cdots,\mathtt{s}_{_{n-1}})$ can be identified with its polynomial representation $\mathtt{s}(z)= \mathtt{s}_{_{0}}+\mathtt{s}_{_{1}}z+\cdots+\mathtt{s}_{_{n-1}}z^{n-1}$ and a cyclic code $\mathtt{C}$ over $\mathtt{R}$ can be observed as an ideal of the quotient ring $\mathtt{R}[z]/\left\langle z^{n}-1\right\rangle.$ A linear code $\mathtt{C}$ is said to be a reversible code if for every $\mathtt{s}=(\mathtt{s}_{_{0}},\mathtt{s}_{_{1}},\cdots,\mathtt{s}_{{n-1}})$ in $\mathtt{C}$, the codeword $\mathtt{s^{r}}= (\mathtt{s}_{_{n-1}},\mathtt{s}_{_{n-2}},\cdots,\mathtt{s}_{_{0}})$ also belongs to $\mathtt{C}$. For a polynomial $g(z)$ of degree $k \leq n-1$, $g^{*}(z)= z^{k}g(z^{-1})$ is defined as its reciprocal polynomial. A polynomial $g(z)$ is said to be self reciprocal if and only if $g^{*}(z)=g(z)$. 

In the following lemmas, we recall some results that are required to proceed further.

\begin{lemma} \label{lemma 1}\cite{jasbir thesis} Let $\mathtt{C}$ be a cyclic code over $\mathtt{R}$ with generators $g_{_{1}}(z),g_{_{2}}(z),\cdots,\\g_{_{k}}(z)$. Then $\mathtt{C}$ is a reversible cyclic code if and only if $g_{_{i}}^{*}(z) \in \mathtt{C}$ for $ 1\leq i\leq k$.
\end{lemma}
\begin{lemma} \label{lemma 2}\cite{jasbir thesis} Let $g_{_{1}}(z), g_{_{2}}(z)$ be any two polynomials in $\mathtt{R}[z]$ with deg $g_{_{1}}(z)\geq$ deg $g_{_{2}}(z)$. Then\\
(i) $\big(g_{_{1}}(z)+g_{_{2}}(z)\big)^{*}= g_{_{1}}^{*}(z)+z^{i}g_{_{2}}^{*}(z)$, where $i= $deg $g_{_{1}}(z)-$ deg $g_{_{2}}(z),$\\
(ii) $\big(g_{_{1}}(z)g_{_{2}}(z)\big)^{*}= g_{_{1}}^{*}(z)g_{_{2}}^{*}(z).$ 
\end{lemma}

\begin{lemma} \label{lemma 3}\cite{rev over Z4} Let $\mathtt{C} = \left\langle g(z)+2p(z), 2a(z) \right\rangle$ be a cyclic code of length $n$ over $Z_{4}$, where $g(z), p(z)$ and $a(z)$ are binary polynomials such that $a(z)|g(z)|z^{n}-1$ and either $p(z)=0$ or $a(z)|p(z)\big(\frac{z^{n}-1}{g(z)}\big)$ with deg $a(z) > $ deg $p(z)$. Then $\mathtt{C}$ is a reversible cyclic code over $Z_{4}$ if and only if\\
(a) $g(z)$ and $a(z)$ are self reciprocal,\\
(b) $a(z)|(z^{\lambda}p^{*}(z)-p(z))$, where $\lambda=$ deg $g(z)-$ deg $p(z) > 0$.\\
\end{lemma}
The rings $Z_{4}+\nu Z_{4}$, $\nu^{2} \in Z_{4}+\nu Z_{4}$ are the extensions of the ring $Z_{4}$ which have been classified into chain rings and non-chain rings by Adel Alahmadi et al. \cite{classification ref}. They have proved that $Z_{4}+\nu Z_{4}$ is a chain ring for $\nu^{2} \in \{2,3,1+\nu,3+\nu,1+2\nu,2+2\nu,1+3\nu,3+3\nu\}$ and is a non-chain ring for $\nu^{2} \in \{0,\nu,2\nu,3\nu,1,3+2\nu,2+\nu,2+3\nu\}.$ Throughout this paper, we will denote the non-chain rings $Z_{4}+\nu Z_{4}, \nu^{2}=\theta$ by $\mathtt{R}_{_\theta}$ for $\theta \in \mathtt{S},$ where $\mathtt{S}= \{0,\nu,2\nu,3\nu,1,3+2\nu,2+\nu,2+3\nu\}.$ 
The complete structure of cyclic codes of arbitrary length over $\mathtt{R}_{_{\theta}},$ $\theta \in \mathtt{S}$ have been established by N. Jain et al.\cite{self paper}. For the sake of completeness we recall the required results. 

Let $\mathtt{C}_{_{\theta}}$ be a cyclic code of length $n$ over $\mathtt{R}_{_{\theta}}$, $\theta \in \mathtt{S}$. Define $\phi_{_{\theta}}$: $\mathtt{R}_{_{\theta}} \rightarrow Z_{4}$ as $\phi_{_{\theta}}(x)=x$ (mod $k_{_{\theta}})$ for $x \in \mathtt{R}_{_{\theta}},$ where
\[ k_{_{\theta}}=
\begin{cases}
	\nu   & ;\theta \in \{0,\nu,2\nu,3\nu\},\\
	1+\nu & ;\theta \in \{1,3+2\nu\},\\
	2+\nu & ;\theta \in \{2+\nu,2+3\nu\}.
\end{cases} \]
The map $\phi_{_{\theta}}$ can be naturally extended to a map $\phi_{_{\theta}}$: $\mathtt{C}_{_{\theta}} \rightarrow Z_{4}[z]/\langle z^{n}-1 \rangle$ as $\phi_{_{\theta}}\big(\mathtt{s}_{_{0}}+\mathtt{s}_{_{1}}z+\cdots+\mathtt{s}_{_{n-1}}z^{n-1}\big) = \phi_{_{\theta}}(\mathtt{s}_{_{0}})+\phi_{_{\theta}}(\mathtt{s}_{_{1}})z+\cdots+\phi_{_{\theta}}(\mathtt{s}_{_{n-1}})z^{n-1}.$ Then, kernel of $\phi_{_{\theta}}$ and torsion of $\mathtt{C}_{_{\theta}}$ are defined as
$ker_{_{\theta}}=\bigl\{a(z) \in \mathtt{C}_{_{\theta}}$ such that $\phi_{_{\theta}}(a(z))=0\bigr\}$ and
Tor($\mathtt{C}_{_{\theta}})$=$\bigl\{b(z)\in \dfrac{Z_{4}[z]}{\langle z^{n}-1\rangle}: k_{_{\theta}} b(z)\in \mathtt{C}_{_{\theta}}\bigr\}$ respectively.\\

\begin{lemma} \label{lemma 4}\cite{self paper} Let $\mathtt{C}_{_{\theta}}$ be a cyclic code of arbitrary length $n$ over the ring $\mathtt{R}_{_{\theta}}, \theta \in \mathtt{S}.$ Then $\mathtt{C}_{_{\theta}}$ is uniquely generated by the polynomials $g_{_{\theta_{1}}}(z),g_{_{\theta_{2}}}(z),g_{_{\theta_{3}}}(z),\\g_{_{\theta_{4}}}(z),$ where $g_{_{\theta_{1}}}(z)= g_{_{11}}(z)+2g_{_{12}}(z)+k_{_{\theta}}g_{_{13}}(z)+2k_{_{\theta}}g_{_{14}}(z)$, $g_{_{\theta_{2}}}(z)=2g_{_{22}}(z)+k_{_{\theta}}g_{_{23}}(z)+2k_{_{\theta}}g_{_{24}}(z)$, $g_{_{\theta_{3}}}(z)=k_{_{\theta}}g_{_{33}}(z)+2k_{_{\theta}}g_{_{34}}(z)$, $g_{_{\theta_{4}}}(z)=2k_{_{\theta}}g_{_{44}}(z)$ such that the polynomials $g_{_{ij}}(z)$ are in $Z_{2}[z]/{\langle z^{n}-1\rangle}$ for $1 \leq i \leq 4, i \leq j \leq 4$  and satisfy the conditions 
\begin{equation*}
	g_{_{22}}(z)|g_{_{11}}(z)|z^{n}-1,~~~~   g_{_{44}}(z)|g_{_{33}}(z)|z^{n}-1,
\end{equation*}
\begin{equation*}
	g_{_{22}}(z)|g_{_{12}}(z)\frac{z^{n}-1}{g_{_{11}}(z)},~~~~ g_{_{44}}(z)|g_{_{34}}(z)\frac{z^{n}-1}{g_{_{33}}(z)}.
\end{equation*}
Also, either $g_{_{ij}}(z)=0$ or deg $g_{_{ij}}(z)<$ deg $g_{_{jj}}(z)$ for $ 1\leq i \leq 3, i<j\leq 4.$ Further, $\phi_{_{\theta}}(\mathtt{C}_{_{\theta}}) =\langle g_{_{11}}(z)+2g_{_{12}}(z),2g_{_{22}}(z)\rangle$ and $ker_{_{\theta}}= k_{_{\theta}}\langle g_{_{33}}(z)+2g_{_{34}}(z),2g_{_{44}}(z)\rangle.$
\end{lemma}
	
	\section{Reversible cyclic codes over $\mathtt{R}_{_{\theta}}$}
In this section, we have shown that the torsion code of a reversible cyclic code over $\mathtt{R}_{_{\theta}},$ $\theta \in \mathtt{S}$ is a reversible cyclic code over $Z_{4}.$ Further, we have obtained sufficient and necessary conditions for a cyclic code $\mathtt{C}_{_{\theta}}$ to be a reversible cyclic code over $\mathtt{R}_{_{\theta}}, \theta \in \mathtt{S}.$
	\begin{theorem} \label{theorem 1}
		Let $\mathtt{C}_{_{\theta}}=\langle g_{_{\theta_{1}}}(z),g_{_{\theta_{2}}}(z),g_{_{\theta_{3}}}(z),g_{_{\theta_{4}}}(z)\rangle$ be a reversible cyclic code of arbitrary length $n$ over the ring $\mathtt{R}_{_{\theta}}, \theta \in \mathtt{S}$, where $g_{_{\theta_{1}}}(z)= g_{_{11}}(z)+2g_{_{12}}(z)+k_{_{\theta}}g_{_{13}}(z)+2k_{_{\theta}}g_{_{14}}(z)$, $g_{_{\theta_{2}}}(z)=2g_{_{22}}(z)+k_{_{\theta}}g_{_{23}}(z)+2k_{_{\theta}}g_{_{24}}(z)$, $g_{_{\theta_{3}}}(z)=k_{_{\theta}}g_{_{33}}(z)+2k_{_{\theta}}g_{_{34}}(z)$, $g_{_{\theta_{4}}}(z)=2k_{_{\theta}}g_{_{44}}(z)$ such that the polynomials $g_{_{ij}}(z)$ are in $Z_{2}[z]/{\left\langle z^{n}-1 \right\rangle}$ for $1\leq i \leq 4, i \leq j \leq 4.$ Then Tor$(\mathtt{C}_{_{\theta}})=\langle g_{_{33}}(z)+2g_{_{34}}(z), 2g_{_{44}}(z)\rangle$ is a reversible cyclic code over $Z_{4}.$
	\end{theorem}
	\begin{proof}
		Let $\mathtt{C}_{_{\theta}}=\langle g_{_{\theta_{1}}}(z),g_{_{\theta_{2}}}(z),g_{_{\theta_{3}}}(z),g_{_{\theta_{4}}}(z)\rangle$ be a reversible cyclic code of arbitrary length $n$ over the ring $\mathtt{R}_{_{\theta}}, \theta \in \mathtt{S}.$ Then from Lemma \ref{lemma 4}, we have $ker_{_{\theta}}$ = $k_{_{\theta}}\langle g_{_{33}}(z)+2g_{_{34}}(z),2g_{_{44}}(z)\rangle.$ Therefore, Tor$(\mathtt{C}_{_{\theta}})=\langle g_{_{33}}(z)+2g_{_{34}}(z),2g_{_{44}}(z)\rangle$. 
		Since $k_{_{\theta}}(g_{_{33}}(z)+2g_{_{34}}(z)) \in \mathtt{C}_{_{\theta}}$ and $\mathtt{C}_{_{\theta}}$ is reversible, therefore, $k_{_{\theta}}(g_{_{33}}(z)+2g_{_{34}}(z))^{*} \in \mathtt{C}_{_{\theta}}$ by Lemma \ref{lemma 1}. It follows that $(g_{_{33}}(z)+2g_{_{34}}(z))^{*} \in$ Tor$(\mathtt{C}_{_{\theta}})$. Similarly, $(2g_{_{44}}(z))^{*} \in$ Tor$(\mathtt{C}_{_{\theta}})$. Hence, Tor$(\mathtt{C}_{_{\theta}})=\langle g_{_{33}}(z)+2g_{_{34}}(z),2g_{_{44}}(z)\rangle$ is a reversible cyclic code over $Z_{4}$ by Lemma \ref{lemma 1}.
			\end{proof}
		The following lemma is easy to prove.
		\begin{lemma} \label{lemma 5} Let $\mathtt{C}_{_{\theta}}$ be a reversible cyclic code of length $n$ over $\mathtt{R}_{_{\theta}}, \theta \in \mathtt{S}$. Then $\phi_{_{\theta}}( \mathtt{C}_{_{\theta}})$ is a reversible cyclic code over $Z_{4}.$ 
	\end{lemma}
	The following theorem gives sufficient and ncessary conditions for a cyclic code $\mathtt{C}_{_{\theta}}$ of an arbitrary length $n$ to be a reversible cyclic code over $\mathtt{R}_{_{\theta}}$.

	\begin{theorem} \label{theorem 2}Let $\mathtt{C}_{_{\theta}}=\langle g_{_{\theta_{1}}}(z),g_{_{\theta_{2}}}(z),g_{_{\theta_{3}}}(z),g_{_{\theta_{4}}}(z)\rangle$ be a cyclic code of arbitrary length $n$ over the ring $\mathtt{R}_{_{\theta}}, \theta \in \mathtt{S}$, where $g_{_{\theta_{1}}}(z)= g_{_{11}}(z)+2g_{_{12}}(z)+k_{_{\theta}}g_{_{13}}(z)+2k_{_{\theta}}g_{_{14}}(z)$, $g_{_{\theta_{2}}}(z)=2g_{_{22}}(z)+k_{_{\theta}}g_{_{23}}(z)+2k_{_{\theta}}g_{_{24}}(z)$, $g_{_{\theta_{3}}}(z)=k_{_{\theta}}g_{_{33}}(z)+2k_{_{\theta}}g_{_{34}}(z)$, $g_{_{\theta_{4}}}(z)=2k_{_{\theta}}g_{_{44}}(z)$ such that the polynomials $g_{_{ij}}(z)$ are in $Z_{2}[z]/{\langle z^{n}-1\rangle}$ for $1\leq i \leq 4, i \leq j \leq 4.$ Also, either $g_{_{ij}}(z)=0$ or deg $g_{_{ij}}(z)<$ deg $g_{_{jj}}(z)$ for $ 1\leq i \leq 3, i<j\leq 4.$ Let $\mathsf{g}_{_{1}}(z)=g_{_{13}}(z)+2g_{_{14}}(z), \mathsf{g}_{_{2}}(z)=g_{_{23}}(z)+2g_{_{24}}(z) \in Z_{4}[z].$ Then $\mathtt{C}_{_{\theta}}$ is a
		reversible cyclic code over $\mathtt{R}_{_{\theta}}$ if and only if
		\begin{itemize}
			\item [$(i)$] $g_{_{ii}}(z), 1\leq i \leq 4,$ are all self reciprocal polynomials,
			\item[$(ii)$] $g_{_{44}}(z)|z^{\alpha}g_{_{34}}^{*}(z)-g_{_{34}}(z),$ where $\alpha=$ deg $g_{_{33}}(z)-$ deg $g_{_{34}}(z) > 0,$
			\item[$(iii)$] $2(z^{\beta}g_{_{12}}^{*}(z)-g_{_{12}}(z)) + k_{_{\theta}}(z^{\gamma} \mathsf{g}_{_{1}}^{*}(z)-\mathsf{g}_{_{1}}(z)) \in \mathtt{C}_{_{\theta}},$ where $\beta =$ deg $g_{_{11}}(z)-$ deg $g_{_{12}}(z) > 0$ and $\gamma =$ deg $g_{_{11}}(z)-$ deg $\mathsf{g}_{_{1}}(z) > 0,$
			\item[$(iv)$] $z^{\delta}\mathsf{g}_{_{2}}^{*}(z)-\mathsf{g}_{_{2}}(z)  \in$ Tor($\mathtt{C}_{_{\theta}}),$ where $\delta =$ deg $g_{_{22}}(z)-$ deg $\mathsf{g}_{_{2}}(z) \geq 0.$ 
		\end{itemize}
	\end{theorem}
	\begin{proof} First, let $\mathtt{C}_{_{\theta}}$ be a reversible cyclic code of length $n$ over $\mathtt{R}_{_{\theta}}$. Then by Lemma \ref{lemma 4} and Lemma \ref{lemma 5}, we have 
		$\phi_{_{\theta}}(\mathtt{C}_{_{\theta}}) = \langle g_{_{11}}(z)+2g_{_{12}}(z), 2g_{_{22}}(z)\rangle$ is a reversible cyclic code over $Z_{4}$. Also, by Theorem \ref{theorem 1}, Tor$(\mathtt{C}_{_{\theta}}) = \langle g_{_{33}}(z)+2g_{_{34}}(z), 2g_{_{44}}(z)\rangle$ is a reversible cyclic code over $Z_{4}$. It follows from Lemma \ref{lemma 3}, $g_{_{11}}(z),g_{_{22}}(z),g_{_{33}}(z)$ and $g_{_{44}}(z)$ are self reciprocal polynomials and $g_{_{44}}(z)|z^{\alpha}g_{_{34}}^{*}(z)-g_{_{34}}(z),$ where $\alpha=$ deg $g_{_{33}}(z)-$ deg $g_{_{34}}(z) > 0.$ This proves conditions $(i)$ and $(ii).$
		As $\mathtt{C}_{_{\theta}}$ is reversible, then by using Lemma \ref{lemma 1}, Lemma \ref{lemma 2} and self reciprocality of $g_{_{11}}(z),$ we have
		$(g_{_{11}}(z)+2g_{_{12}}(z)+k_{_{\theta}}\mathsf{g}_{_{1}}(z))^{*}= g_{_{11}}^{*}(z)+2z^{\beta}g_{_{12}}^{*}(z)+k_{_{\theta}} z^{\gamma}\mathsf{g}_{_{1}}^{*}(z)= g_{_{11}}(z)+2z^{\beta}g_{_{12}}^{*}(z)+k_{_{\theta}} z^{\gamma}\mathsf{g}_{_{1}}^{*}(z)  \in \mathtt{C}_{_{\theta}},$ where $\beta=$ deg $g_{_{11}}(z)-$ deg $g_{_{12}}(z) > 0$ and $\gamma =$ deg $g_{_{11}}(z)-$ deg $\mathsf{g}_{_{1}}(z) > 0$. 
		This implies that, 
		\begin{equation}\label{equation 1}
			\begin{split}
				&g_{_{11}}(z)+2z^{\beta}g_{_{12}}^{*}(z)+k_{_{\theta}} z^{\gamma}\mathsf{g}_{_{1}}^{*}(z)= A(z)\big (g_{_{11}}(z)+2g_{_{12}}(z)+k_{_{\theta}}\mathsf{g}_{_{1}}(z)\big)
				+B(z)\big(2g_{_{22}}(z)+\\&k_{_{\theta}} \mathsf{g}_{_{2}}(z)\big) 
				+ k_{_{\theta}} C(z)\big(g_{_{33}}(z)+2g_{_{34}}(z)\big)+k_{_{\theta}} D(z)\big(2g_{_{44}}(z)\big)
			\end{split}
		\end{equation}
		for some $A(z),B(z),C(z),D(z) \in \mathtt{R}_{_{\theta}}[z]$. Multiplying equation (\ref{equation 1}) by $2k_{_{\theta}}$ for $\theta \in \{0,1,2\nu,3+2\nu\}$ and by $2(k_{_{\theta}}-1)$ for $\theta \in \{\nu,3\nu,2+\nu,2+3\nu\}$ on both sides, we get 
		\begin{equation}\label{equation 2}
			\begin{cases}
				2k_{_{\theta}} g_{_{11}}(z)= 2k_{_{\theta}} A(z)g_{_{11}}(z) & \text{for } \theta \in \{0,1,2\nu,3+2\nu\}\\
				2(k_{_{\theta}}-1) g_{_{11}}(z)= 2(k_{_{\theta}}-1) A(z)g_{_{11}}(z)  &\text{for } \theta \in \{\nu,3\nu,2+\nu,2+3\nu\}\\
			\end{cases}
		\end{equation} 
		Comparing the degrees on both sides of equation (\ref{equation 2}), we find that $A(z)$ is constant. Further it is observed that $A(z)=1+2a+k_{_{\theta}}b,$ where $a \in Z_{2}$ and $b \in Z_{4}.$
		Putting the value of $A(z)$ in equation (\ref{equation 1}) we get,
		$2z^{\beta}g_{_{12}}^{*}(z)+k_{_{\theta}} z^{\gamma}\mathsf{g}_{_{1}}^{*}(z) = 2g_{_{12}}(z)+k_{_{\theta}}\mathsf{g}_{_{1}}(z)+(2a+k_{_{\theta}}b)\big(g_{_{11}}(z)+2g_{_{12}}(z)+k_{_{\theta}}\mathsf{g}_{_{1}}(z)\big)+B(z)\big(2g_{_{22}}(z)+k_{_{\theta}} \mathsf{g}_{_{2}}(z)\big)
		+ k_{_{\theta}} C(z)\big(g_{_{33}}(z)+2g_{_{34}}(z)\big)+k_{_{\theta}} D(z)\big(2g_{_{44}}(z)\big),$ which implies that
		$2\big(z^{\beta}g_{_{12}}^{*}(z)-g_{_{12}}(z)\big)+k_{_{\theta}}\big( z^{\gamma}\mathsf{g}_{_{1}}^{*}(z)-\mathsf{g}_{_{1}}(z)\big)=(2a+k_{_{\theta}}b)\big(g_{_{11}}(z)+2g_{_{12}}(z)+k_{_{\theta}}\mathsf{g}_{_{1}}(z)\big)+B(z)\big(2g_{_{22}}(z)+k_{_{\theta}} \mathsf{g}_{_{2}}(z)\big)
		+ k_{_{\theta}} C(z)\big(g_{_{33}}(z)+2g_{_{34}}(z)\big)+k_{_{\theta}} D(z)\big(2g_{_{44}}(z)\big) \in \mathtt{C}_{_{\theta}}.$ It follows that
		$2\big(z^{\beta}g_{_{12}}^{*}(z)-g_{_{12}}(z)\big)+k_{_{\theta}}\big( z^{\gamma}\mathsf{g}_{_{1}}^{*}(z)-\mathsf{g}_{_{1}}(z)\big) \in \mathtt{C}_{_{\theta}},$
		which proves condition $(iii).$
		As $\mathtt{C}_{_{\theta}}$ is reversible, using Lemma \ref{lemma 1}, Lemma \ref{lemma 2} and self reciprocality of $g_{_{22}}(z),$ we have $\big(2g_{_{22}}(z)+k_{_{\theta}} \mathsf{g}_{_{2}}(z)\big)^{*}=2g_{_{22}}^{*}(z) + k_{_{\theta}} z^{\delta}\mathsf{g}_{_{2}}^{*}(z)=2g_{_{22}}(z) +k_{_{\theta}} z^{\delta}\mathsf{g}_{_{2}}^{*}(z) \in \mathtt{C}_{_{\theta}},$ where $\delta =$ deg $g_{_{22}}(z) -$ deg $\mathsf{g}_{_{2}}(z) \geq 0$. This implies that $2g_{_{22}}(z) + k_{_{\theta}} \mathsf{g}_{_{2}}(z) +k_{_{\theta}}\big(z^{\delta}\mathsf{g}_{_{2}}^{*}(z)-\mathsf{g}_{_{2}}(z)\big) \in \mathtt{C}_{_{\theta}}.$ As $2g_{_{22}}(z) + k_{_{\theta}} \mathsf{g}_{_{2}}(z) \in \mathtt{C}_{_{\theta}},$ it follows that $k_{_{\theta}}\big(z^{\delta}\mathsf{g}_{_{2}}^{*}(z)-\mathsf{g}_{_{2}}(z)\big) \in \mathtt{C}_{_{\theta}}$.   Thus, $z^{\delta}\mathsf{g}_{_{2}}^{*}(z)-\mathsf{g}_{_{2}}(z) \in$ Tor($\mathtt{C}_{_{\theta}}).$
		
		Conversely, suppose all the conditions $(i)-(iv)$ hold. In order to prove that $\mathtt{C}_{_{\theta}}$ is a reversible cyclic code over $\mathtt{R}_{_{\theta}}$, from Lemma \ref{lemma 1}, it is enough to show that
		$(g_{_{11}}(z)+2g_{_{12}}(z)+k_{_{\theta}} \mathsf{g}_{_{1}}(z))^{*},(2g_{_{22}}(z)+k_{_{\theta}} \mathsf{g}_{_{2}}(z))^{*},k_{_{\theta}} (g_{_{33}}(z)+2g_{_{34}}(z))^{*}$ and $2k_{{_\theta}}(g_{_{44}}(z))^{*} \in \mathtt{C}_{_{\theta}}.$
		Using Lemma \ref{lemma 2} and condition $(i)$ we have,
		$(g_{_{11}}(z)+2g_{_{12}}(z)+k_{_{\theta}} \mathsf{g}_{_{1}}(z))^{*}
		=g_{_{11}}^{*}(z)+2z^{\beta}g_{_{12}}^{*}(z)+k_{_{\theta}} z^{\gamma}\mathsf{g}_{_{1}}^{*}(z)
		=(g_{_{11}}(z)+2g_{_{12}}(z)+k_{_{\theta}} \mathsf{g}_{_{1}}(z)) +2(z^{\beta}g_{_{12}}^{*}(z)-g_{_{12}}(z))
		+k_{_{\theta}}(z^{\gamma}\mathsf{g}_{_{1}}^{*}(z)- \mathsf{g}_{_{1}}(z))$ which belongs to $\mathtt{C}_{_{\theta}}$, by condition $(iii).$
		Again using Lemma \ref{lemma 2} and condition $(i)$ we have,
		$(2g_{_{22}}(z)+k_{_{\theta}} \mathsf{g}_{_{2}}(z))^{*}=2g_{_{22}}^{*}(z)+k_{_{\theta}} z^{\delta}\mathsf{g}_{_{2}}^{*}(z)=(2g_{_{22}}(z)+k_{_{\theta}} \mathsf{g}_{_{2}}(z))+k_{_{\theta}} (z^{\delta}\mathsf{g}_{_{2}}^{*}(z)- \mathsf{g}_{_{2}}(z))$
		which belongs to $\mathtt{C}_{_{\theta}}$, by condition $(iv).$
		Similarly, using Lemma \ref{lemma 2}, condition $(i)$ and $(ii)$ we have 
		$k_{_{\theta}} (g_{_{33}}(z)+2g_{_{34}}(z))^{*}
		=k_{_{\theta}}(g_{_{33}}^{*}(z)+ 2z^{\alpha}g_{_{34}}^{*}(z))
		=k_{_{\theta}}(g_{_{33}}(z)+2g_{_{34}}(z))+2k_{_{\theta}}(z^{\alpha}g_{_{34}}^{*}(z)- g_{_{34}}(z))
		=k_{_{\theta}}(g_{_{33}}(z)+2g_{_{34}}(z))+2k_{_{\theta}} s(z)g_{_{44}}(z) \text{ for some}~ s(z) \in Z_{4}[z].$ It clearly belongs to $\mathtt{C}_{_{\theta}}.$
		Finally, $2k_{_{\theta}}(g_{_{44}}(z))^{*}= 2k_{_{\theta}} g_{_{44}}(z)$ belongs to $\mathtt{C}_{_{\theta}}$ by condition $(i).$
		\end{proof}
	
	Following examples act as an illustration of our results.
	
	\begin{example} \label{example 1}Let $\mathtt{C}_{_{\theta}} = \langle z^{3}+z^{2}+z+1, 2(z^{2}+1)+2\nu, \nu(z^{2}+1), 2\nu (z+1) \rangle$ be a cyclic code of length $4$ over the ring $\mathtt{R}_{_{\theta}}$ for $\theta=2\nu.$ Here $g_{_{11}}(z)=z^{3}+z^{2}+z+1, g_{_{22}}(z)=z^{2}+1,  g_{_{33}}(z)=z^{2}+1, g_{_{44}}(z)=z+1, g_{_{12}}(z)=0, g_{_{34}}(z)=0, \mathsf{g}_{_{1}}(z)=0 ,\mathsf{g}_{_{2}}(z)=2.$ Then we have,
	\begin{itemize}
		\item [(i)] $g_{_{11}}^{*}(z)=1+z+z^{2}+z^{3}, g_{_{22}}^{*}(z)=1+z^{2}, g_{_{33}}^{*}(z)=1+z^{2}, g_{_{44}}^{*}(z)=1+z.$
		\text{Thus}, $g_{_{ii}}^{*}(z)=g_{_{ii}}(z)$ for $1 \leq i \leq 4.$
		
		\item[(ii)] Since $\alpha=2$ \text{ and} $g_{_{34}}(z)=0,$ \text{it implies that}  $g_{_{44}}(z)|z^{2}g_{_{34}}^{*}(z)-g_{_{34}}(z).$
		
		\item[(iii)] Since $\beta=\gamma=3$ \text{and} $g_{_{12}}(z)=\mathsf{g}_{_{1}}(z)=0,$ \text{which implies that} $2\big(z^{3} g_{_{12}}^{*}(z)-g_{_{12}}(z)\big)+\nu \big(z^{3}\mathsf{g}_{_{1}}^{*}(z)-\mathsf{g}_{_{1}}(z)\big)=0 \in \mathtt{C}_{_\theta}.$ 
		
		\item[(iv)] Since $\delta =2$ \text{and} $\mathsf{g}_{_{2}}^{*}(z)=2,$ \text{thus} $z^{2}\mathsf{g}_{_{2}}^{*}(z)-\mathsf{g}_{_{2}}(z)= 2z^{2}-2= (z-1)\big(2(z+1)\big) \in$ Tor($\mathtt{C}_{_{\theta}}).$ 
	\end{itemize}   
	Hence, $\mathtt{C}_{_{\theta}}$ is a reversible cyclic code as it satisfies all the conditions of Theorem \ref{theorem 2}.
\end{example}
	
	\begin{example} Let $\mathtt{C}_{_{\theta}} = \langle z^{4}+z^{3}+z+1, 2(z^{2}+z+1)+(2+\nu)(z^{2}+z+1), (2+\nu)(z^{4}+z^{3}+z+1), 2(2+\nu)(z^{2}+z+1) \rangle$ be a cyclic code of length $6$ over the ring $\mathtt{R}_{_{\theta}}$ for $\theta=2+\nu.$ Here $g_{_{11}}(z)=z^{4}+z^{3}+z+1, g_{_{22}}(z)=z^{2}+z+1,  g_{_{33}}(z)=z^{4}+z^{3}+z+1, g_{_{44}}(z)=z^{2}+z+1, g_{_{12}}(z)=0, g_{_{34}}(z)=0, \mathsf{g}_{_{1}}(z)=0 ,\mathsf{g}_{_{2}}(z)=z^{2}+z+1.$ We have,
	\begin{itemize}
		\item [(i)] $g_{_{11}}^{*}(z)=1+z+z^{3}+z^{4}, g_{_{22}}^{*}(z)=1+z+z^{2}, g_{_{33}}^{*}(z)=1+z+z^{3}+z^{4}, g_{_{44}}^{*}(z)=1+z+z^{2}.$
		\text{So}, $g_{_{ii}}^{*}(z)=g_{_{ii}}(z)$ for $1 \leq i \leq 4.$
		
		\item[(ii)] Since $\alpha=4 \text{ and } g_{_{34}}(z)=0,$ \text{it implies that}  $g_{_{44}}(z)|z^{4}g_{_{34}}^{*}(z)-g_{_{34}}(z).$
		
		\item[(iii)] As $\beta=\gamma =4, g_{_{12}}(z)=\mathsf{g}_{_{1}}(z)=0,$ \text{ we see that} $2\big(z^{4} g_{_{12}}^{*}(z)-g_{_{12}}(z)\big)+(2+\nu) \big(z^{4}\mathsf{g}_{_{1}}^{*}(z)-\mathsf{g}_{_{1}}(z)\big)=0 \in \mathtt{C}_{_\theta}.$ 
		
		\item[(iv)] As $\delta= 0$ and $\mathsf{g}_{_{2}}^{*}(z)=z^{2}+z+1,$ \text{thus we have} $z^{0}\mathsf{g}_{_{2}}^{*}(z)-\mathsf{g}_{_{2}}(z)=0 \in$ Tor($\mathtt{C}_{_{\theta}}).$ 
	\end{itemize}   
	Hence, $\mathtt{C}_{_{\theta}}$ satisfies all the conditions of Theorem \ref{theorem 2}. Therefore, $\mathtt{C}_{_{\theta}}$ is a reversible cyclic code.
\end{example}
	\begin{example} Let $\mathtt{C}_{_{\theta}} = \langle z^{3}+z^{2}+z+1+(1+\nu), 2(z^{2}+1),(1+\nu)(z-1), 2(1+\nu)\rangle$ be a cyclic code of length $4$ over the ring $\mathtt{R}_{_{\theta}}$ for $\theta=3+2\nu.$ Here $g_{_{11}}(z)=z^{3}+z^{2}+z+1, g_{_{22}}(z)=z^{2}+1,  g_{_{33}}(z)=z-1, g_{_{44}}(z)=1, g_{_{12}}(z)=0, g_{_{34}}(z)=0, \mathsf{g}_{_{1}}(z)=1 ,\mathsf{g}_{_{2}}(z)=0.$ We have,
	\begin{itemize}
		\item [(i)] $g_{_{11}}^{*}(z)=1+z+z^{2}+z^{3}, g_{_{22}}^{*}(z)=1+z^{2}, g_{_{33}}^{*}(z)=1+z, g_{_{44}}^{*}(z)=1.$
		\text{So}, $g_{_{ii}}^{*}(z)=g_{_{ii}}(z)$ for $1 \leq i \leq 4.$
		
		\item[(ii)] Since $\alpha=1$ and $g_{_{34}}(z)=0,$ \text{it implies that}  $g_{_{44}}(z)|zg_{_{34}}^{*}(z)-g_{_{34}}(z).$
		
		\item[(iii)] Since $\beta=\gamma=3, g_{_{12}}(z)=0$ and $\mathsf{g}_{_{1}}(z)=1,$ \text{we see that} $2\big(z^{3} g_{_{12}}^{*}(z)-g_{_{12}}(z)\big)+(1+\nu) \big(z^{3}\mathsf{g}_{_{1}}^{*}(z)-\mathsf{g}_{_{1}}(z)\big)= (1+\nu)(z^{3}-1)=(z^{2}+z+1)\big((1+\nu)(z-1)\big) \in \mathtt{C}_{_\theta}.$ 
		
		\item[(iv)] As $\delta=2$ and $\mathsf{g}_{_{2}}^{*}(z)=0,$ \text{thus} $z^{2}\mathsf{g}_{_{2}}^{*}(z)-\mathsf{g}_{_{2}}(z)=0 \in$ Tor($\mathtt{C}_{_{\theta}}).$ 
	\end{itemize}   
	Hence, $\mathtt{C}_{_{\theta}}$ satisfies all the conditions of Theorem \ref{theorem 2}. Therefore, $\mathtt{C}_{_{\theta}}$ is a reversible cyclic code. 
\end{example}
	\begin{example} Let $\mathtt{C}_{_{\theta}} = \langle z^{3}+z^{2}+z+1+\nu(z+3), 2(z^{2}+1)+2\nu, \nu(z^{2}+1), 2\nu(z+1) \rangle$ be a cyclic code of length $4$ over the ring $\mathtt{R}_{_{\theta}}$ for $\theta =2\nu.$ Here $g_{_{11}}(z)=z^{3}+z^{2}+z+1, g_{_{22}}(z)=z^{2}+1,  g_{_{33}}(z)=z^{2}+1, g_{_{44}}(z)=z+1, g_{_{12}}(z)=0, g_{_{34}}(z)=0, \mathsf{g}_{_{1}}(z)=z+3 ,\mathsf{g}_{_{2}}(z)=2.$ We have,
	\begin{itemize}
		\item [(i)] $g_{_{11}}^{*}(z)=1+z+z^{2}+z^{3},~  g_{_{22}}^{*}(z)=1+z^{2},~ g_{_{33}}^{*}(z)=1+z^{2},~ g_{_{44}}^{*}(z)=1+z.$
		\text{So}, $g_{_{ii}}^{*}(z)=g_{_{ii}}(z)$ for $1 \leq i \leq 4.$
		
		\item[(ii)] Since $\alpha=2$ and $g_{_{34}}(z)=0,$ \text{it implies that}  $g_{_{44}}(z)|z^{2}g_{_{34}}^{*}(z)-g_{_{34}}(z).$
		
		\item[(iii)] As $\beta=3,\gamma=2, g_{_{12}}(z)=0$ and $\mathsf{g}_{_{1}}(z)=z+3,$ \text{ which implies that} $2\big(z^{3} g_{_{12}}^{*}(z)-g_{_{12}}(z)\big)+\nu \big(z^{2}\mathsf{g}_{_{1}}^{*}(z)-\mathsf{g}_{_{1}}(z)\big)= \nu(3z^{3}+z^{2}+3z+1)=(3z+1)\big(\nu(z^{2}+1)\big) \in \mathtt{C}_{_\theta}.$ 
		
		\item[(iv)] As $\delta=2$ and $\mathsf{g}_{_{2}}^{*}(z)=2,$ \text{thus} $z^{2}\mathsf{g}_{_{2}}^{*}(z)-\mathsf{g}_{_{2}}(z)=2z^{2}+2=2\big(z^{2}+1\big) \in$ Tor($\mathtt{C}_{_{\theta}}).$ 
	\end{itemize}   
	Hence, $\mathtt{C}_{_{\theta}}$ satisfies all the conditions of Theorem \ref{theorem 2}. Therefore, $\mathtt{C}_{_{\theta}}$ is a reversible cyclic code. 
\end{example}
	
	\begin{example} Let $\mathtt{C}_{_{\theta}} = \langle z^{5}+z^{4}+z^{3}+z^{2}+z+1+\nu(z^{4}+z^{2}+1), 2(z+1)+\nu(z+1), \nu(z^{5}+z^{4}+z^{3}+z^{2}+z+1), 2\nu \rangle$ be a cyclic code of length $6$ over the ring $\mathtt{R}_{_{\theta}}$ for $\theta =\nu.$ Here $g_{_{11}}(z)=z^{5}+z^{4}+z^{3}+z^{2}+z+1, g_{_{22}}(z)=z+1,  g_{_{33}}(z)=z^{5}+z^{4}+z^{3}z^{2}+z+1, g_{_{44}}(z)=1, g_{_{12}}(z)=0, g_{_{34}}(z)=0, \mathsf{g}_{_{1}}(z)=z^{4}+z^{2}+1 ,\mathsf{g}_{_{2}}(z)=z+1.$ We have,
	\begin{itemize}
		\item [(i)] $g_{_{11}}^{*}(z)=z^{5}+z^{4}+z^{3}+z^{2}+z+1,~  g_{_{22}}^{*}(z)=1+z,~ g_{_{33}}^{*}(z)=z^{5}+z^{4}+z^{3}+z^{2}+z+1,~ g_{_{44}}^{*}(z)=1.$
		\text{So}, $g_{_{ii}}^{*}(z)=g_{_{ii}}(z)$ for $1 \leq i \leq 4.$
		
		\item[(ii)] Since $\alpha=2$ and $g_{_{34}}(z)=0,$ \text{it implies that}  $g_{_{44}}(z)|z^{2}g_{_{34}}^{*}(z)-g_{_{34}}(z).$
		
		\item[(iii)] As $\beta=5,\gamma=1, g_{_{12}}(z)=0$ and $\mathsf{g}_{_{1}}(z)=z^{4}+z^{2}+1,$ \text{ which implies that} $2\big(z^{5} g_{_{12}}^{*}(z)-g_{_{12}}(z)\big)+\nu \big(z\mathsf{g}_{_{1}}^{*}(z)-\mathsf{g}_{_{1}}(z)\big)= \nu(z^{5}+3z^{4}+z^{3}+3z^{2}+z+3)=\nu\big(z^{5}+z^{4}+z^{3}+z^{2}+z+1\big)+2\nu(z^{4}+z^{2}+1) \in \mathtt{C}_{_\theta}.$ 
		
		\item[(iv)] As $\delta=0$ and $\mathsf{g}_{_{2}}^{*}(z)=z+1,$ \text{thus} $z^{0}\mathsf{g}_{_{2}}^{*}(z)-\mathsf{g}_{_{2}}(z)=0 \in$ Tor($\mathtt{C}_{_{\theta}}).$ 
	\end{itemize}   
	Hence, $\mathtt{C}_{_{\theta}}$ satisfies all the conditions of Theorem \ref{theorem 2}. Therefore, $\mathtt{C}_{_{\theta}}$ is a reversible cyclic code. 
\end{example}
	
	\begin{example} \label{example 6} Let $\mathtt{C}_{_{\theta}} = \langle z^{5}+z^{4}+z^{3}+z^{2}+z+1+\nu(z^{2}+z+1)+2\nu z, 2(z^{4}+z^{2}+1), \nu(z^{3}+3), 2\nu(z^{2}+z+1) \rangle$ be a cyclic code of length $6$ over the ring $\mathtt{R}_{_{\theta}}$ for $\theta =0.$ Here $g_{_{11}}(z)=z^{5}+z^{4}+z^{3}+z^{2}+z+1, g_{_{22}}(z)=z^{4}+z^{2}+1,  g_{_{33}}(z)=z^{3}+3, g_{_{44}}(z)=z^{2}+z+1, g_{_{12}}(z)=0, g_{_{34}}(z)=0, \mathsf{g}_{_{1}}(z)=z^{2}+3z+1 ,\mathsf{g}_{_{2}}(z)=0.$ Clearly, $g_{_{33}}^{*}(z)=3z^{3}+1 \ne g_{_{33}}(z)$ which violates condition $(i)$ of Theorem \ref{theorem 2}. Hence, $\mathtt{C}_{_{\theta}}$ is not a reversible cyclic code.
\end{example}
	\section{Reversible complement cyclic codes over $\mathtt{R}_{_{\theta}}$}
	In this section, we obtain conditions for a cyclic code of arbitrary length over $\mathtt{R}_{_{\theta}}, \theta \in \mathtt{S}$ to be a reversible complement cyclic code which are necessary as well as sufficient.
	For this, we shall use the generalized notion of complement of an element over a finite commutative ring given by J. Kaur et al. \cite{jasbir DAM}.
	
	\begin{definition}\cite{jasbir DAM} For an element $a \in \mathtt{R}_{_{\theta}}, \overline{a}$ is known as the complement of $a$ with respect to $u_{_{\theta}}$ and $t_{_{\theta}}$ if $a+u_{_{\theta}}\overline{a}=t_{_{\theta}},$ where $u_{_{\theta}}$ is a unit in $\mathtt{R}_{_{\theta}}$ and $t_{_{\theta}}$ is an arbitrary element of $\mathtt{R}_{_{\theta}}$ such that $u^{2}_{_{\theta}}=1$ and $u_{_{\theta}}t_{_{\theta}}=t_{_{\theta}}.$ We shall denote the complement of $a$ with respect to $u_{_{\theta}}$ and $t_{_{\theta}}$ by $(\overline{a})_{(u_{_{\theta}},t_{_{\theta}})}.$ 
\end{definition}
	
	\begin{definition} A cyclic code $\mathtt{C}_{_{\theta}}$ of length $n$ over $\mathtt{R}_{_{\theta}}$ is called a $(u_{_{\theta}},t_{_{\theta}})$ reversible complement cyclic code if $\big((\overline{\mathtt{s}}_{_{n-1}})_{(u_{_{\theta}},t_{_{\theta}})},(\overline{\mathtt{s}}_{_{n-2}})_{(u_{_{\theta}},t_{_{\theta}})},\cdots,(\overline{\mathtt{s}}_{_{0}})_{(u_{_{\theta}},t_{_{\theta}})}\big) \in \mathtt{C}_{_{\theta}},$ whenever $(\mathtt{s}_{_{0}},\mathtt{s}_{_{1}},\cdots,\mathtt{s}_{_{n-1}}) \in \mathtt{C}_{_{\theta}}.$
	\end{definition}
	
	\begin{definition} For a polynomial $\mathtt{s}(z)$ of degree $k \leq n-1,$ its reverse polynomial is $\mathtt{s}^{\mathtt{r}}(z)=z^{n-k-1}s^{*}(z).$
\end{definition}
	
	\begin{definition} The $(u_{_{\theta}},t_{_{\theta}})$ reverse complement of the polynomial representation $\mathtt{s}(z)$ of the codeword $(\mathtt{s}_{_{0}},\mathtt{s}_{_{1}},\cdots,\mathtt{s}_{_{n-1}})$ is the polynomial representation of the element $\big((\overline{\mathtt{s}}_{_{n-1}})_{(u_{_{\theta}},t_{_{\theta}})},(\overline{\mathtt{s}}_{_{n-2}})_{(u_{_{\theta}},t_{_{\theta}})},\cdots,(\overline{\mathtt{s}}_{_{0}})_{(u_{_{\theta}},t_{_{\theta}})}\big).$ We shall denote  $(\overline{\mathtt{s^{\mathtt{r}}}(z)})_{(u_{_{\theta}},t_{_{\theta}})}.$ 
\end{definition}

The following lemma follows easily from the definition of the complement.

\begin{lemma} For any $r_{_{1}},r_{_{2}},r_{_{3}} \in \mathtt{R}_{_{\theta}},$ we have
	\begin{enumerate}
		\item $\Big(\overline{(\overline{r_{_{1}}})_{(u_{_{\theta}},t_{_{\theta}})}}\
		\Big)_{(u_{_{\theta}},t_{_{\theta}})}=r_{_{1}}.$
		\item $(\overline{r_{_{1}}+r_{_{2}}})_{(u_{_{\theta}},t_{_{\theta}})}=(\overline{r_{_{1}}})_{(u_{_{\theta}},t_{_{\theta}})}+(\overline{r_{_{2}}})_{(u_{_{\theta}},t_{_{\theta}})}+3u_{_{\theta}}^{-1}t_{_{\theta}}.$
		\item $(\overline{r_{_{1}}+t_{_{\theta}}r_{_{2}}})_{(u_{_{\theta}},t_{_{\theta}})}=(\overline{r_{_{1}}})_{(u_{_{\theta}},t_{_{\theta}})}+3u_{_{\theta}}^{-1}t_{_{\theta}}r_{_{2}}.$
		\item $u_{_{\theta}}(\overline{r_{_{1}}})_{(u_{_{\theta}},t_{_{\theta}})}+3t_{_{\theta}}=3r_{_{1}}.$
		\item $(\overline{r_{_{1}}+r_{_{2}}+r_{_{3}}})_{(u_{_{\theta}},t_{_{\theta}})}=(\overline{r_{_{1}}})_{(u_{_{\theta}},t_{_{\theta}})}+(\overline{r_{_{2}}})_{(u_{_{\theta}},t_{_{\theta}})}+(\overline{r_{_{3}}})_{(u_{_{\theta}},t_{_{\theta}})}+2u_{_{\theta}}^{-1}t_{_{\theta}}.$
	\end{enumerate}
\end{lemma}

The following theorem gives conditions for a cyclic code $\mathtt{C}_{_{\theta}}$ of an arbitrary length $n$ over $\mathtt{R}_{_{\theta}}$ to be a $(u_{_{\theta}},t_{_{\theta}})$ reversible complement cyclic code  which are necessary as well as sufficient.
	\begin{theorem} Let $ u_{_{\theta}},t_{_{\theta}} \in \mathtt{R}_{_{\theta}},$ such that $u_{_{\theta}}$ is a unit. A cyclic code $\mathtt{C}_{_{\theta}}$ of length $n$ over $\mathtt{R}_{_{\theta}}$ is a $(u_{_{\theta}},t_{_{\theta}})$ reversible complement cyclic code if and only if $\mathtt{C}_{_{\theta}}$ is a reversible cyclic code and $(\overline{0^{\mathtt{r}}(z)})_{(u_{_{\theta}},t_{_{\theta}})}\in \mathtt{C}_{_{\theta}}.$ 
	\end{theorem}
	\begin{proof}Firstly, suppose that $\mathtt{C}_{_{\theta}}$ is a reversible cyclic code of length $n$ over $\mathtt{R}_{_{\theta}}$ and $(\overline{0^{\mathtt{r}}(z)})_{(u_{_{\theta}},t_{_{\theta}})}\in \mathtt{C}_{_{\theta}}.$ Let $\mathtt{s}(z)=\mathtt{s}_{_{0}}+\mathtt{s}_{_{1}}z+\cdots+\mathtt{s}_{_{k}}z^{k}; 0\leq k \leq n-1$ be an arbitrary polynomial in $\mathtt{C}_{_{\theta}}.$ Since $\mathtt{C}_{_{\theta}}$ is a reversible cyclic code, therefore $\mathtt{s}^{\mathtt{r}}(z)=z^{n-k-1}\mathtt{s}^{*}(z)= \mathtt{s}_{_{k}}z^{n-k-1}+\mathtt{s}_{_{k-1}}z^{n-k}+\cdots+\mathtt{s}_{_{0}}z^{n-1} \in \mathtt{C}_{_{\theta}}.$ Also $(\overline{0^{\mathtt{r}}(z)})_{(u_{_{\theta}},t_{_{\theta}})}\in \mathtt{C}_{_{\theta}}.$ Thus, $(\overline{0^{\mathtt{r}}(z)})_{(u_{_{\theta}},t_{_{\theta}})}-u_{_{\theta}}^{-1}\mathtt{s}^{\mathtt{r}}(z) \in \mathtt{C}_{_{\theta}}.$ Moreover,
		\begin{align*}
			&(\overline{0^{\mathtt{r}}(z)})_{(u_{_{\theta}},t_{_{\theta}})}-u_{_{\theta}}^{-1}\mathtt{s}^{\mathtt{r}}(z)=u_{_{\theta}}^{-1}t_{_{\theta}}(1+z+z^{2}+\cdots+z^{n-1})-u_{_{\theta}}^{-1}(\mathtt{s}_{_{k}}z^{n-k-1}+\\& \mathtt{s}_{_{k-1}}z^{n-k}+\cdots+\mathtt{s}_{_{0}}z^{n-1})
			=u_{_{\theta}}^{-1}t_{_{\theta}}(1+z+z^{2}+\cdots+z^{n-k-2})+\big(u_{_{\theta}}^{-1}(t_{_{\theta}}-\mathtt{s}_{_{k}})\\&z^{n-k-1}+u_{_{\theta}}^{-1}(t_{_{\theta}}-\mathtt{s}_{_{k-1}})z^{n-k}+\cdots+u_{_{\theta}}^{-1}(t_{_{\theta}}-\mathtt{s}_{_{0}})z^{n-1}\big)
			=u_{_{\theta}}^{-1}t_{_{\theta}}(1+z+z^{2}+\\&\cdots+z^{n-k-2})+\big((\overline{\mathtt{s}}_{_{k}})_{(u_{_{\theta}},t_{_{\theta}})}z^{n-k-1}+(\overline{\mathtt{s}}_{_{k-1}})_{(u_{_{\theta}},t_{_{\theta}})}z^{n-k}+\cdots+(\overline{\mathtt{s}}_{_{0}})_{(u_{_{\theta}},t_{_{\theta}})}z^{n-1}\big)\\&
			=(\overline{\mathtt{s}^{\mathtt{r}}(z)})_{(u_{_{\theta}},t_{_{\theta}})},
		\end{align*}
		i.e., for each polynomial $\mathtt{s}(z) \in \mathtt{C}_{_{\theta}},  (\overline{\mathtt{s}^{\mathtt{r}}(z)})_{(u_{_{\theta}},t_{_{\theta}})} \in \mathtt{C}_{_{\theta}}.$ Hence, $\mathtt{C}_{_{\theta}}$ is a $(u_{_{\theta}},t_{_{\theta}})$ reversible complement cyclic code.\\
		
		Conversely, suppose that $\mathtt{C}_{_{\theta}}$ is a $(u_{_{\theta}},t_{_{\theta}})$ reversible complement cyclic code. Let  $\mathtt{s}(z)=\mathtt{s}_{_{0}}+\mathtt{s}_{_{1}}z+\cdots+\mathtt{s}_{_{k}}z^{k}; 0\leq k \leq n-1$ be an arbitrary polynomial in $\mathtt{C}_{_{\theta}}.$ Since  $\mathtt{C}_{_{\theta}}$ is a $(u_{_{\theta}},t_{_{\theta}})$ reversible complement cyclic code, therefore  $(\overline{\mathtt{s}^{\mathtt{r}}(z)})_{(u_{_{\theta}},t_{_{\theta}})} \in \mathtt{C}_{_{\theta}}.$ In particular, $(\overline{0^{\mathtt{r}}(z)})_{(u_{_{\theta}},t_{_{\theta}})} \in \mathtt{C}_{_{\theta}}.$ Therefore, $(\overline{0^{\mathtt{r}}(z)})_{(u_{_{\theta}},t_{_{\theta}})}-(\overline{\mathtt{s}^{\mathtt{r}}(z)})_{(u_{_{\theta}},t_{_{\theta}})}=u_{_{\theta}}^{-1}\mathtt{s}^{r}(z)=u_{_{\theta}}^{-1}z^{n-k-1}s^{*}(z) \in \mathtt{C}_{_{\theta}}.$ It follows that, $s^{*}(z) \in \mathtt{C}_{_{\theta}}, \mathtt{C}_{_{\theta}}$ being a cyclic code. Thus, $\mathtt{C}_{_{\theta}}$ is a reversible cyclic code over $\mathtt{R}_{_{\theta}}.$
	\end{proof}
	In the following table, we have checked the cyclic codes from Example \ref{example 1}- Example \ref{example 6} whether they are $(u_{_{\theta}},t_{_{\theta}})$ reversible complement cyclic codes or not for some values of  $u_{_{\theta}}$ and $t_{_{\theta}}.$
\begin{center}
	\begin{tabular}{|c|c|c|c|c|c|}
		\hline
		\textbf{S.No.} & \textbf{Cyclic code $\mathtt{C}{_{_\theta}}$}  & $\theta$ & \textbf{Reversible}  & $(u_{_{\theta}},t_{_{\theta}})$ & \multicolumn{1}{m{2cm}|}{\textbf{Reversible Complement}}\\
		\hline 
		
		1 & \multicolumn{1}{m{5cm}|}{$ \langle z^{3}+z^{2}+z+1,$  $2(z^{2}+1)+2\nu,\nu(z^{2}+1), 2\nu (z+1)\rangle$} & $2\nu$ &  Yes & all possible values  & Yes \\
		\hline
		2 & \multicolumn{1}{m{5cm}|}{$\langle z^{4}+z^{3}+z+1,2(z^{2}+z+1)+(2+\nu)(z^{2}+z+1), (2+\nu)(z^{4}+z^{3}+z+1),2(2+\nu)(z^{2}+z+1)\rangle$} & $2+\nu$ & Yes & $(1,2+\nu)$ & Yes \\
		\hline
		3 & \multicolumn{1}{m{5cm}|}{$\langle z^{4}+z^{3}+z+1,2(z^{2}+z+1)+(2+\nu)(z^{2}+z+1), (2+\nu)(z^{4}+z^{3}+z+1),2(2+\nu)(z^{2}+z+1)\rangle$} & $2+\nu$ & Yes & $(1+2\nu,2\nu)$ & No \\
		\hline
		4 & \multicolumn{1}{m{5cm}|}{$\langle z^{3}+z^{2}+z+1+(1+\nu),2(z^{2}+1),(1+\nu)(z-1),2(1+\nu) \rangle$} & $3+2\nu$ & Yes & $(3+2\nu,2)$ & Yes\\
		\hline
		5 & \multicolumn{1}{m{5cm}|}{$\langle z^{3}+z^{2}+z+1+(1+\nu),2(z^{2}+1),(1+\nu)(z-1),2(1+\nu)\rangle$} & $3+2\nu$ &Yes & $(3,2+2\nu)$ & No\\ 
		\hline
		6 & \multicolumn{1}{m{5cm}|}{$\langle z^{3}+z^{2}+z+1+\nu(z+3),2(z^{2}+1)+2\nu,\nu(z^{2}+1),2\nu(z+1)\rangle$} & $2\nu$ & Yes & $(1,\nu)$ & Yes\\
		\hline
		7 & \multicolumn{1}{m{5cm}|}{$\langle z^{3}+z^{2}+z+1+\nu(z+3),2(z^{2}+1)+2\nu,\nu(z^{2}+1),2\nu(z+1)\rangle$} & $2\nu$ & Yes & $(1,3+\nu)$ & No\\ 
		\hline
		8 & \multicolumn{1}{m{5cm}|}{$\langle z^{5}+z^{4}+z^{3}+z^{2}+z+1+\nu(z^{4}+z^{2}+1),2(z+1)+\nu(z+1), \nu(z^{5}+z^{4}+z^{3}+z^{2}+z+1),2\nu\rangle$} & $\nu$ & Yes & $(1,\nu)$ & Yes \\ 
		\hline
		9 & \multicolumn{1}{m{5cm}|}{$\langle z^{5}+z^{4}+z^{3}+z^{2}+z+1+\nu(z^{4}+z^{2}+1),2(z+1)+\nu(z+1), \nu(z^{5}+z^{4}+z^{3}+z^{2}+z+1),2\nu\rangle$} & $\nu$ & Yes & $(1+2\nu,2\nu)$ & No\\ 
		\hline
		10 & \multicolumn{1}{m{5cm}|}{$\langle z^{5}+z^{4}+z^{3}+z^{2}+z+1+\nu(z^{2}+z+1)+2\nu z,2(z^{4}+z^{2}+1), \nu(z^{3}+3),2\nu(z^{2}+z+1)\rangle$} & $0$ & No & all possible values & No \\
		\hline
		
	\end{tabular}
\end{center}
	\section{Conclusion} In this paper, sufficient and necessary conditions for a cyclic code of arbitrary length over the ring $Z_{4}+\nu Z_{4}$ to be a reversible cyclic code have been established for those values of $\nu^{2}$ for which $Z_{4}+\nu Z_{4}$ is a non-chain ring. Also, conditions for a cyclic code over these rings to be a reversible complement cyclic code which are necessary as well sufficient have been determined .

\end{document}